\documentclass[10 pt]{article}
\usepackage{epsfig}
\begin{document}

\title{\bf Speeding up the Grover algorithm using auxiliary solutions}

\author{A.Y. Shiekh\footnote{\rm shiekh@dinecollege.edu} \\
             {\it Din\'{e} College, Tsaile, Arizona, U.S.A.}}

\date{}

\maketitle

\abstract{It may be possible to extend the Grover search algorithm by taking a divide and conquer approach using auxiliary solutions to achieve an exponential speed-up.}

\begin{flushright}
``{\it What is proved by impossibility proofs is lack of imagination}'' \\
John Bell
\end{flushright}

\baselineskip .65 cm

\section{Introduction}
To date there is only one generic search algorithm for quantum computers \cite{NielsenChuang}, namely the Grover algorithm, and that achieves a quadratic speed-up over a classical computer.

It is noted here, that the Grover approach is particularly efficient if one quarter of the candidate solutions are valid, and a proposal is made to try and take advantage of this observation by adding auxiliary solutions.

\section{Grover's algorithm reviewed}
For the sake of illustration we look at the $n$ bit case. Starting with ${\it \left| 0 \right>}$ one applies the Hadamard transformation to get the usual spread of ($N = 2^n$) candidate solutions
\begin{equation}
\left(
{\it \left| 0 \right>} +
\left| 1 \right> +
\cdots +
\left| N-1 \right>
\right)/\sqrt{N}
\end{equation}
${\it \left| 0 \right>}$ will have a rather special role to play in this approach, and unlike the others, does not represent a candidate solution\footnote{${\it \left| 0 \right>}$ can be separately checked to determine if it is a solution or not, and regardless can be marked as invalid in the processing that follows.}.

One then applies the Grover four step procedure (to be repeated as many times as necessary)

\begin{itemize}

\item Mark the $M$ valid solution(s) (represented by ${\bf \left| S_n \right>}$) with a $\pi$ phase change (which is unitary and so allowed). The resultant state is given by:
\begin{equation}
\left( 
\begin{array}{rrrrr}
{\it \left| 0 \right>} +
\left| 1 \right> +
\cdots +
&+ {\bf \left| S_1 \right>}& +
\cdots 
&+ {\bf \left| S_M \right>}& +
\cdots +
\left| N - 1 \right>
\\
&-2 {\bf \left| S_1 \right>}& +
\cdots
&-2 {\bf \left| S_M \right>}&
\end{array} 
\right)/\sqrt{N}
\end{equation}

\item Perform a Hadamard transformation ($H$) that will undo the first line to yield:
\begin{equation}
\begin{array}{c}
{\it \left| 0 \right>}
\\
- 2 \frac{1}{\sqrt{N}} H( {\bf \left| S_1 \right>} + \cdots + {\bf \left| S_M \right>})
\end{array} 
\end{equation}

\item Perform a $\pi$ phase change on all but the ${\it \left| 0 \right>}$ state
\begin{equation}
\begin{array}{c}
{\it \left| 0 \right>} - 4\frac{M}{N} {\it \left| 0 \right>}
\\
+2 \frac{1}{\sqrt{N}} H ( {\bf \left| S_1 \right>} + \cdots+ {\bf \left| S_M \right>})
\end{array} 
\end{equation}
Crucial to this manipulation is the fact that the Hadamard transform of any candidate begins with $+ {\it {\it \left| 0 \right>}}/\sqrt{N}$.

\item Perform another Hadamard transformation that will restore the initial Hadamard spread, but with a reduced amplitude.
\begin{equation}
\begin{array}{rrl}
(1 - 4 \frac{M}{N})
(
{\it \left| 0 \right>} +
\left| 1 \right> +
\cdots +
&{\bf \left| S_1 \right>} +
\cdots +
{\bf \left| S_M \right>}& +
\cdots +
\left| N-1 \right>
)/\sqrt{N}
\\
+2
(
&{\bf \left| S_1 \right>} +
\cdots +
{\bf \left| S_M \right>}&
)/\sqrt{N}
\end{array} 
\label{eqn:iteration}
\end{equation}

\end{itemize}

What has been achieved is a lowering of the invalid solution amplitudes, and a compensatory lifting of the valid solution amplitudes. In summary, after a single iteration, the probability of finding an invalid solution is given by
\begin{equation}
(N-M)\left(\frac{1-4M/N}{\sqrt{N}}\right)^2
\end{equation}
and that of finding a valid solution is
\begin{equation}
M\left(\frac{1-4M/N}{\sqrt{N}} + \frac{2}{\sqrt{N}} \right)^2
\label{eqn:valid}
\end{equation}
and together they indeed sum to unity as they must.

Of particular note is the observation that if one quarter of the candidates are valid, only a single run is needed to eliminate all the invalid solutions, and it is hoped to take advantage of this to speed up the Grover approach.

However, before we do that, it would be good to recover the performance of the standard Grover approach. Begin by defining the normalized sums of valid and invalid solutions: $\left| \alpha \right> \equiv (\sum invalid)/\sqrt{N-M}$ and $ \left| \beta \right> \equiv (\sum valid)/\sqrt{M}$; then the starting state $\left| \psi \right>$ is given by
\begin{equation}
\left| \psi \right> =
\sqrt{\frac{N-M}{N}} \left| \alpha \right> + \sqrt{\frac{M}{N}} \left| \beta \right>
\end{equation}
and the state after one Grover application $G$ is given (from equation \ref{eqn:iteration}) by
\begin{equation}
G \left| \psi \right> =
\left(1 - 4 \frac{M}{N} \right) \sqrt{\frac{N-M}{N}} \left| \alpha \right> 
+ 
\left(1 - 4 \frac{M}{N} + 2 \right) \sqrt{\frac{M}{N}} \left| \beta \right>
\end{equation}
Re-expressing these in terms of the fraction of valid solutions $f \equiv M/N$, gives these as
\begin{equation}
\left| \psi \right> =
\sqrt{1-f} \left| \alpha \right> + \sqrt{f} \left| \beta \right>
\end{equation}
\begin{equation}
G \left| \psi \right> =
(1-4f) \sqrt{1-f} \left| \alpha \right> + (3-4f) \sqrt{f} \left| \beta \right>
\end{equation}
Then expressing $\sqrt{f}$ as $\sin{\theta/2}$ and using the two trigonometric identities $\cos{3\phi} = (1 - 4 \sin^2 \phi) \cos \phi$ and $\sin{3\phi} = (3 - 4 \sin^2 \phi) \sin \phi$ simplifies these to
\begin{equation}
\left| \psi \right> =
\cos \frac{\theta}{2} \left| \alpha \right> + \sin \frac{\theta}{2} \left| \beta \right>
\end{equation}
\begin{equation}
G \left| \psi \right> =
\cos \frac{3\theta}{2} \left| \alpha \right> + \sin \frac{3\theta}{2} \left| \beta \right>
\end{equation}
yielding the picture that each Grover application rotates the state toward the valid solution set by an angle $\theta$ given by $\sin{\theta/2} = \sqrt{M/N}$; so it takes the usual Grover approach order $\sqrt{N/M}$ iterations to expose the valid solution set.

Having reviewed the Grover algorithm, we now return to the observation that if one quarter of the solutions are valid, they are isolated in a single iteration of the Grover approach.

\section{Divide and conquer mechanism using auxiliary solutions}
What is significant about the Grover algorithm is that, up to a point, the more valid solutions that are present the faster it runs, and it is hoped to take advantage of this mechanism by adding (and later removing) auxiliary solutions.

\begin{figure}[ht]
   \centering
   \includegraphics[width = 3.5in]{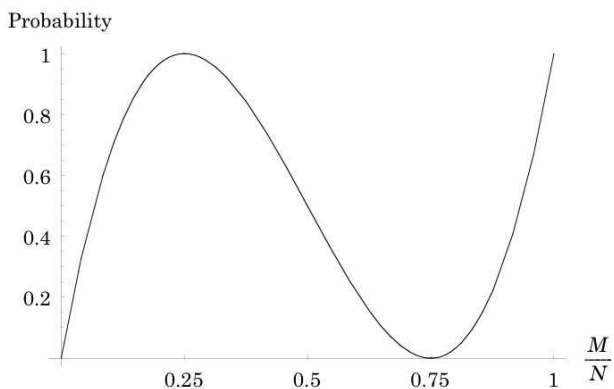}
   \caption{Probability versus fraction of valid solutions}
   \label{fig:graph}
\end{figure}

The probability of finding a valid solution after just one iteration is given by equation \ref{eqn:valid} to be
\begin{equation}
f (3-4f)^2
\end{equation}
where $f \equiv M/N$ (the fraction of valid solutions), and as noted before, an optimum is reached when a quarter of the solutions are valid, at which point all invalid solutions are removed in a single iteration of the Grover procedure (see figure \ref{fig:graph}).

The following supplementary procedure (to be repeated as many times as necessary) is suggested: `Add' additional auxiliary solutions by considering a candidate solution to be valid not only if it satisfies the criterion to be solved for, but also if the first two bits are each $1$. In this way very close to a quarter of the states will be valid\footnote{This could be adjusted to be exactly one quarter if the number of valid solutions were known beforehand.}, and a single run will then eliminate the invalid three quarters.

Having eliminated all invalid solutions not beginning with $1 1$ one would like to repeat the procedure, but only on the remaining quarter.

What one now has (from equation \ref{eqn:iteration}) is
\begin{equation}
\frac{1}{\sqrt{M+N_\epsilon}}
\left(
{\bf \left| S_1 \right>} +
\cdots +
{\bf \left| S_M \right>}
+ \sum \left| \epsilon \right>
\right)
\end{equation}
having used $M = N/4$, and where $\left| \epsilon \right>$ represents any true solution not beginning with $1 1$ that might have been present in the three quarters that was up for rejection. 

Now, ${\bf \left| S_1 \right>} + \cdots + {\bf \left| S_M \right>}$ is the set of all combinations beginning with $1 1$, so the above may be rewritten as:
\begin{equation}
\frac{1}{\sqrt{M+N_\epsilon}}
\left(
\left| 1 1 \right>
\left(
{\it \left| 0 \right>}_{-2} +
\left| 1 \right>_{-2} +
\cdots +
\left| M-1 \right>_{-2}
\right)
+ \sum \left| \epsilon \right>
\right)
\end{equation}
where the new Hadamard spread is now in the space two digits smaller. One might then repeat the Grover approach, using the relevant reduced Hadamard transformation, but not changing the marking function that continues to look at all digits. Each iteration cuts the number of remaining solutions in quarter, so achieving exponential performance.

At completion one will be left with the true solutions as well as the state with all ones, namely $\left| N-1 \right>$ and this would need to be checked separately, as was also the case for ${\it \left| 0 \right>}$, albeit for different reasons.

\section{Conclusion}

It may be possible to speed up the Grover approach beyond quadratic, by exploiting its somewhat counter-intuitive feature of improved performance, up till the point where one quarter of the solutions are valid. One quarter valid solutions is artificially achieved by temporarily recognizing one in four of the candidate solutions as valid.

Another aspect of this approach is that it uses only the digital aspects of quantum theory for calculation, which might greatly simplify the error correction procedure. The problem with analogue systems is that errors, all be them small, creep into {\em all} aspects of the system in any finite time, and so are impossible to remove completely. In contrast, digital systems have the advantage that the probability of an error is generally small, albeit that the error itself, if seen, is large; so for a small time interval, it is very unlikely that all aspects of the system find themselves in error.

This proposal gets around the optimality proof  as the `Oracle' changes for each iteration.

\end{document}